\documentclass[12pt]{article}
\usepackage[english]{babel}
\usepackage{graphicx}
\usepackage{graphics}
\usepackage{exscale}
\usepackage{epsfig}
\usepackage{amsmath}
\begin{document}
\begin{center}
{\LARGE{\bf Electric  polarizabilities of proton and neutron  and the
relativistic  center-of-mass coordinate}}
\\[1ex]
R.N. Lee$^a$, A.I. Milstein$^a$, M. Schumacher$^b$\\
{\em
$^a$Budker Institute of Nuclear Physics, 630090 Novosibirsk, Russia\\
$^b$Zweites Physikalisches Institut der Universit\"at
G\"ottingen,\\
D-37073 G\"ottingen, Germany\\}
\end{center}
\begin{abstract}
We argue that the relativistic correction $\delta{\bf R}_{c.m.}$
to the center-of-mass vector can lead to the approximate equality
of the proton and neutron electric polarizabilities in the quark
model. The explicit form of $\delta{\bf R}_{c.m.}$ depends only on
the non-relativistic potential between quarks. In particular, this
correction is the same for the potential generated by
Lorentz-vector and -scalar interactions.
\end{abstract}

\section{Introduction}

The electric polarizability $\bar\alpha$ and magnetic
polarizability $\bar\beta$  are fundamental structure constants of
the nucleon having a direct relation to the internal dynamics of the
particle. Their magnitudes depend not only on the quantum numbers
of the constituents, but also on the properties of the interaction
between these constituents. The  prediction of the electromagnetic
polarizabilities  and the comparison with experimental data may
serve as a sensitive tool for tests of hadron models.

 The quantities  $\bar\alpha$ and  $\bar\beta$ can be obtained from the
amplitude for low-energy Compton scattering  off the nucleon. In the lab
frame, this
amplitude up to $O(\omega^2)$ terms reads
\cite{klein55,petrunkin61}

\begin{equation}\label{T}
T=T_{Born}+\bar\alpha\omega_1\omega_2{\boldsymbol \epsilon}_1\cdot
{\boldsymbol \epsilon}_2^*+\bar\beta
({\bf k}_1\times{\boldsymbol \epsilon}_1)\cdot({\bf k}_2\times
{\boldsymbol \epsilon}_2^*)\ ,
\end{equation}
where $\omega_i,\, {\bf k}_i,$ and ${\boldsymbol \epsilon}_i$ are
the energy, momentum, and polarization vector of the incoming ($i=1$)
and outgoing ($i=2$) photons ($\hbar=c=1$). The contribution
$T_{Born}$ corresponds to the amplitude for  Compton scattering off
a  point-like particle with the spin, mass, charge, and magnetic
moment of the nucleon.

 The result of the most recent and most precise  experimental
investigation \cite{olmos01,kossert02}, obtained by
 Compton scattering off the proton below  pion threshold,
is shown in the first line of Table \ref{table}. For the neutron
the same method is extremely difficult \cite{rose90}: (i)
Quasi-free Compton scattering from neutrons bound in the deuteron
has to be carried out leading to sizable corrections at low
energies. (ii) The Born amplitude $T_{Born}$ is very small for the
neutron. Therefore,  the corresponding interference term is also
small and, hence,  cannot be used as is  done in case of the
proton. (iii) The low-energy differential cross section is less
sensitive to the electric polarizability in case of the neutron as
compared to the proton.

In order to avoid these difficulties, the
differential cross section for electromagnetic scattering of
neutrons in the Coulomb field of heavy nuclei \cite{lvov93}
\begin{equation}
\frac{d\sigma_{\rm pol}}{d\Omega}= \pi M p (Z e)^2 {\rm Re}\,a
\left\{\bar \alpha_n \sin\frac{\theta}{2} - \frac{e^2 \kappa^2_n}{2 M^3}
\left(1-\sin\frac{\theta}{2}\right)\right\}
\label{coul}
\end{equation}
has been investigated in a series of narrow-beam neutron
transmission  experiments. In (\ref{coul}) $p$ is the neutron
momentum, $-a$  the nuclear amplitude, $\kappa_n$ the neutron
anomalous magnetic moment, $M$ the neutron mass, and $e$  the
proton charge. Of these experiments only the Oak Ridge experiment
\cite{schmiedmayer91} has a reasonable precision.
This result is cited in the second line of Table \ref{table} where  the
necessary correction for the second (Schwinger) term in the braces
of (\ref{coul}) has been carried out.  This Schwinger-term correction was
disregarded in the original evaluation of the experiment
\cite{schmiedmayer91} and reads $+e^2 \kappa^2_n/4 M^3= +0.62$ \cite{lvov93}.

In \cite{levchuk94} it was shown that it is also possible to use
experimental data on quasi-free Compton scattering on neutrons
bound in deuterons at energies between $\pi$ meson threshold and
the $\Delta$ peak to extract the electric
polarizability of the neutron. This method has successfully been
tested for the proton  \cite{wissmann99} and applied to the neutron
\cite{kossert02,kolb00}, leading to the result listed in the third
line of Table \ref{table}.
\begin{center}
\begin{table}[h]
{\caption{Experimental electric polarizabilities for proton and
neutron}}\vspace{1mm} \label{table}
\centering\begin{tabular}{|l|l|l|} \hline $\bar\alpha_p$&$12.2\pm
0.3({\rm stat})\pm 0.4({\rm syst})\pm 0.3({\rm model})$&
Compton\cite{olmos01,kossert02}\\
$\bar\alpha_n$&$12.6\pm 1.5({\rm stat})\pm 2.0({\rm
  syst})$&Coulomb\cite{schmiedmayer91}\\
$\bar\alpha_n$&$12.5\pm 1.8({\rm stat}){}^{+1.1}_{-0.6}({\rm
syst})
\pm 1.1({\rm model})$&Compton\cite{kossert02}\\
\hline
\end{tabular}
\end{table}
\end{center}
From  Table \ref{table} we obtain the following result for the
difference between the electric polarizabilities of proton and
neutron
\begin{equation}
\bar\alpha_p - \bar\alpha_n = -0.3 \pm 1.8 \label{difference}
\end{equation}
which apparently is compatible with zero.

The electromagnetic polarizabilities of hadrons have been calculated
in many different models.
Though much effort  has been devoted to these  model calculations,
all of them can not be considered as completely satisfactory. In
particular, there is a problem in explaining within
a non-relativistic quark model that the electric polarizabilities
of proton and neutron are equal to each other as suggested in
(\ref{difference}).
It was derived many years ago
\cite{petrunkin61,baldin60,petrunkin64} that $\bar \alpha$ can be
represented as a sum
\begin{equation} \label{alpha}
\bar \alpha = \frac23\sum_{k \neq 0} \frac{|\langle k|{\bf D}|0
\rangle|^2} {E_k-E_0}+ \Delta\alpha
 = \alpha_\circ + \Delta\alpha\, ,
\end{equation}
where $\bf D$ is the internal electric dipole operator,
$|0\rangle$ and $|k\rangle$ are the ground and excited states in
terms of internal coordinates,  $E_k$ and $E_0$ are the
corresponding energies. Note that ${\bf P}|0\rangle={\bf
P}|k\rangle=0$, where
 ${\bf P}$ is the operator of total momentum of the particle.
The term $\Delta\alpha$ in $\bar \alpha$
has a relativistic origin  and its leading term is
equal to
\begin{equation} \label{deltaalpha}
 \Delta\alpha=\frac{Z\, e^2\,r_E^2}{3M}\, ,
\end{equation}
where $Z$ and $M$ are the particle electric charge number and
mass, and $r_E$ is the electric radius defined through the Sachs
form factor $G_E$. The quantity $\alpha_\circ$ calculated in a
non-relativistic quark model without taking into account
relativistic corrections leads to the same magnitude   of
$\alpha_\circ$ for  proton and neutron.  Since $\Delta\alpha$ is
equal to zero for the neutron but gives a significant contribution
to $\bar\alpha$ for the proton, one can naively expect that
\begin{equation}
\bar{\alpha}_p -\bar{\alpha}_n =\Delta\alpha_p= 3.8 \pm0.1 \, .
\label{diffbar}
\end{equation}
Comparing (\ref{difference}) with (\ref{diffbar}) we would come to
the conclusion that the experimental and predicted differences
between the electric polarizabilities of proton and neutron
deviate from each other by $2.3\sigma$.   In the following it will
be shown that this discrepancy may be connected with hitherto
unknown relativistic corrections to $\alpha_\circ$ which are of
the same order as $\Delta\alpha$. The relativistic corrections to
$\alpha_\circ$  in (\ref{alpha}) come from  corrections to the
wave functions, to the energies of the ground and excited states,
and a correction to $\bf D$. The relativistic correction to the
electric  dipole operator is connected with the appropriate
relativistic definition of the center-of-mass coordinate
\cite{clos71,friar77,lmsh01}. The neglect of this correction leads
to an incomplete expression for $\bar \alpha$, and, as it has been
shown in \cite{lmsh01}, the missing  piece can be very essential.

In the present paper we show that the difference
$\delta\alpha=(\alpha_\circ)_p-(\alpha_\circ)_n$ is due to the
correction to the electric  dipole operator only. By carrying out
model calculations we show that $\delta\alpha$ can strongly
compensate $\Delta\alpha$, thus removing the contradiction between
(\ref{difference}) and  (\ref{diffbar}) mentioned above.

\section{Correction to the center-of-mass vector}

Let us consider a system of spin-1/2 particles having the masses
 $m_i$ and spin operators ${\bf s}_i$. The Hamiltonian of the system $H$ ,
 accounting for the first relativistic correction, is the sum of the non-relativistic
Hamiltonian  $H_{nr}$ and the Breit Hamiltonian $H_B$, where
\begin{eqnarray}
H_{nr}=\sum_{i} \frac{ p^2_i}{2m_i}+\sum_{i>j}V_{ij}(r_{ij})\, .
\end{eqnarray}
Here ${\bf r}_{ij}={\bf r}_i- {\bf r}_j$ and $V_{ij}(r)$ are some
potentials.  The explicit form of the Breit Hamiltonian $H_B$
depends on the Lorentz structure of interaction, namely,  whether
it is   Lorentz-scalar or Lorentz-vector (see \cite{olsmil83} and
references therein). As in  most models, we assume that the
potential $V(r)$ contains a short-range Coulomb-type term $V_C(r)$
due to  single-gluon exchange corresponding to vector interaction,
and a long-range confining potential $V_{conf}(r)$. The Lorentz
structure of the latter is not well known. We show that it makes
no difference in the form of $\delta{\bf R}_{c.m.}$.

The center-of-mass vector ${\bf R}_{c.m.}$  should satisfy the
following relations \cite{clos71}:
\begin{eqnarray}
[ R_{c.m.}^j\, ,\,  P^k]&=&i\delta^{jk}\, ,\nonumber\\
{}[{\bf R}_{c.m.}\, ,\, H_{nr}+H_B]&=& i\frac{\bf P}{M+H}\approx
i\frac{\bf P}{M} \left(1-\frac{H_{nr}}{M}\right)\, ,
\end{eqnarray}
where $M$ and ${\bf P}$ are the total mass and the total momentum
of the system, respectively. Using the explicit form of the  Breit
Hamiltonian it is easy to check that in both cases, i.e.
Lorentz-scalar and Lorentz-vector interactions, the vector ${\bf
R}_{c.m.}$ for the system of particles has the form
\begin{eqnarray}\label{rcmgeneral}
{\bf R}_{c.m.}&=&{\bf R}+\delta{\bf R}_0+\delta{\bf R}_s\, ,
\quad {\bf R}=\frac{1}{M}\sum_{i}m_i{\bf r}_i \, ,\nonumber\\
\delta {\bf R}_0&=&\frac{1}{2M}\sum_{i}\left\{{\bf r}_i-{\bf R}\,
, \, \frac{ p^2_i}{2m_i}+\frac{1}{2}\sum_{j\ne i}V_{ij}({\bf r}_i-
{\bf r}_j)
\right\}  \, ,\nonumber\\
\delta {\bf R}_s&=&\frac{1}{2M}\sum_{i}\left(\frac{ {\bf p}_i}{m_i}
\times  {\bf s}_i \right )\, ,
\end{eqnarray}
where $M=\sum m_i$, and where the notation $\{a,b\}=ab+ba$ is
used. Thus, the relativistic correction $\delta{\bf R}_0$ is
expressed via the non-relativistic potential of the interaction
between the constituents.

Let us consider proton and neutron. In this case all potentials
are equal to each other, i.e.  $V_{ij}=V$. Let the vectors ${\bf r}_1$
and ${\bf r}_2$ correspond to two $u$-quarks in the proton and two
$d$-quarks in the neutron,  and ${\bf r}_3$ correspond to the
$d$-quark in the proton and the $u$-quark in the neutron, and $m_1$ to the
mass of the identical quarks ($u$ quarks in the proton and $d$-quarks
in the neutron), and $m_3$ to  the mass of the third quark. We pass to
the Jacobi variables:
\begin{eqnarray}\label{Jacobi}
{\boldsymbol\rho}=\frac{1}{\sqrt 2}({\bf r}_1- {\bf r}_2)\quad ,
\quad {\boldsymbol\lambda}=\frac{1}{\sqrt 6}({\bf r}_1+ {\bf
r}_2-2{\bf r}_3)\, , \quad {\bf R}=\frac{1}{M}(m_1{\bf r}_1+m_1
{\bf r}_2+m_3{\bf r}_3)\, ,
\end{eqnarray}
where $M=2m_1+m_3$. Then the  momentum operators are
\begin{eqnarray}
{\bf p}_{\rho}=\frac{1}{\sqrt 2}({\bf p}_1- {\bf p}_2)\, , \quad
{\bf p}_{\lambda}=\frac{\sqrt 6}{2M}(m_3{\bf p}_1+m_3 {\bf p}_2-
2m_1{\bf p}_3)\, , \quad {\bf P}={\bf p}_1+{\bf p}_2+{\bf
p}_3\quad .
\end{eqnarray}
The non-relativistic Hamiltonian $H_{nr}$ in terms of the Jacobi
variables has the form
\begin{eqnarray}\label{HnrJacobi}
H_{nr}= \frac{ P^2}{2M}+\frac{ p^2_{\rho}}{2m_1}+ \frac{
p^2_{\lambda}}{2m_{\lambda}}+V(\sqrt{2}{\rho})+
V(\sqrt{2}{\xi})+V(\sqrt{2}{\eta}) \, ,
\end{eqnarray}
where $m_{\lambda}=3m_1m_3/M$,
${\boldsymbol\xi}=({\boldsymbol\rho}+
\sqrt{3}{\boldsymbol\lambda})/2$, and
${\boldsymbol\eta}=({-\boldsymbol\rho}+
\sqrt{3}{\boldsymbol\lambda})/2$. In terms of Jacobi variables the
spin-independent correction to the center-of-mass vector is
\begin{eqnarray}\label{rcm0difm}
&&\delta {\bf  R}_0= \frac{1}{2M^2}\left(\{{\boldsymbol\rho}\, ,\,
{\bf P}\cdot{\bf p}_\rho\}+\left\{{\boldsymbol\lambda}\, ,\, {\bf
P}\cdot {\bf p}_\lambda\right\}\right)
\nonumber \\
&&+\frac{1}{2\sqrt{6}m_1M}\{{\boldsymbol\rho}\, ,\, {\bf
p}_\lambda\cdot{\bf p}_\rho\}+
\frac{\sqrt{6}}{8M^2}\left\{{\boldsymbol\lambda}\, ,\,
\frac{m_3}{m_1}
p_\rho^2+\frac{m_3-2m_1}{m_\lambda}p_\lambda^2\right\}
\nonumber \\
&&+{\boldsymbol\lambda}\frac{\sqrt{6}m_3V(\sqrt{2}{\rho})}{2M^2}+
\left[\frac{\boldsymbol\rho}{\sqrt 2}+
\frac{\sqrt{6}(m_3-2m_1){\boldsymbol\lambda}}{2M}\right]
\frac{V(\sqrt{2}{\xi})}{2M}\nonumber\\
&&+\left[-\frac{\boldsymbol\rho}{\sqrt 2}+
\frac{\sqrt{6}(m_3-2m_1){\boldsymbol\lambda}}{2M}\right]
\frac{V(\sqrt{2}{\eta})}{2M}\ ,
\end{eqnarray}
and the spin-dependent correction
\begin{eqnarray}\label{rcmsdifm}
\delta {\bf  R}_s=\frac{{\bf P}\times {\bf S}}{2M^2}+ \frac{{\bf
p}_\rho\times ({\bf s}_1-{\bf s}_2)}{2\sqrt{2}m_1M}+ \frac{{\bf
p}_\lambda\times [m_3({\bf s}_1+{\bf s}_2)-2m_1{\bf s}_3]}
{2\sqrt{6}m_1m_3M} \ .
\end{eqnarray}

When $m_1=m_3=m$ , then $m_\lambda=m$ and the expressions for
$\delta{\bf R}_0$ and $\delta{\bf R}_s$ become essentially
simpler:
\begin{eqnarray}\label{rcm0eqm}
&&\delta {\bf R}_0=\frac{1}{18m^2}\left[\{{\boldsymbol\rho}\, ,\,
{\bf P}\cdot{\bf p}_\rho\}+ \left\{{\boldsymbol\lambda}\, ,\, {\bf
P}\cdot {\bf p}_\lambda\right\}\right]
\nonumber \\
&&+\frac{1}{6\sqrt{6}m^2}\left[\{{\boldsymbol\rho}\, ,\, {\bf
p}_\lambda\cdot{\bf p}_\rho\}+
\frac{1}{2}\left\{{\boldsymbol\lambda}\, ,\,
p_\rho^2-p_\lambda^2\right\}\right]
\nonumber \\
&&+\frac{1}{6\sqrt{6}m}\left[
2{\boldsymbol\lambda}V(\sqrt{2}{\rho})+ \left({\sqrt
3}{\boldsymbol\rho}-{\boldsymbol\lambda}\right)V(\sqrt{2}{\xi})
+\left(-{\sqrt 3}{\boldsymbol\rho}-{\boldsymbol\lambda}\right)
V(\sqrt{2}{\eta})\right]\ ,
\end{eqnarray}
and
\begin{eqnarray}\label{rcmseqm}
\delta {\bf R}_s=\frac{1}{6m^2}\left[\frac{1}{3}{\bf P}\times {\bf
S}+ \frac{1}{\sqrt{2}}{\bf p}_\rho\times ({\bf s}_1-{\bf s}_2)+
\frac{1}{\sqrt{6}}{\bf p}_\lambda\times ({\bf s}_1+{\bf s}_2-2{\bf
s}_3)\right] \ .
\end{eqnarray}

Let us consider the substitutions
\begin{eqnarray}\label{subs}
&& {\boldsymbol\rho}\to
\frac{1}{2}{\boldsymbol\rho}-\frac{\sqrt{3}}{2}{\boldsymbol\lambda}
\, \quad ,\quad
{\boldsymbol\lambda}\to\frac{\sqrt{3}}{2}{\boldsymbol\rho}+
\frac{1}{2}{\boldsymbol\lambda} \, ,\nonumber\\
&& {\bf p}_\rho\to\frac{1}{2}{\bf p}_\rho -\frac{\sqrt{3}}{2}{\bf
p}_\lambda  \quad ,\quad {\bf p}_\lambda\to\frac{\sqrt{3}}{2}{\bf
p}_\rho+ \frac{1}{2}{\bf p}_\lambda \, .
\end{eqnarray}
If $m_1=m_3$ , then the non-relativistic Hamiltonian
(\ref{HnrJacobi}) is invariant under this substitution and
$\delta{\bf R}_0({\bf P}=0)\to -\delta{\bf R}_0({\bf P}=0)$. In
addition, the Hamiltonian (\ref{HnrJacobi}) is invariant under the
transformation ${\boldsymbol\rho}\to -{\boldsymbol\rho}$ and
${\boldsymbol\lambda}\to -{\boldsymbol\lambda}$. The properties of
the operators with respect to the transformations (\ref{subs}) are
useful for the selection rules for the matrix elements.

\section{Charge radii}

Let us consider  the charge radii of the nucleon $r^2_E=\langle\,
\sum e_i({\bf r}_i-{\bf R}_{c.m.})^2\,\rangle $, where $e_i$ is
equal to $2/3$ for the $u$-quark and $-1/3$ for the $d$-quark:
 \begin{eqnarray}\label{radii}
(r^2_E)_p&=&\frac{1}{3}\langle 0|\left[4({\bf r}_1-{\bf
R}_{c.m.})^2-
({\bf r}_3-{\bf R}_{c.m.})^2\right]|0\rangle\, , \nonumber \\
(r^2_E)_n&=& -\frac{2}{3}\langle 0|\left[({\bf r}_1-{\bf
R}_{c.m.})^2- ({\bf r}_3-{\bf R}_{c.m.})^2\right]|0\rangle\, .
\end{eqnarray}
Here we used the symmetry of the wave function with respect to the
permutation ${\bf r}_1\leftrightarrow  {\bf r}_2$.
The values of $(r^2_E)_p$ and $(r^2_E)_n$ were measured to be
 $(r^2_E)_p=0.74\,\mbox{fm}^2$ and  $(r^2_E)_n=-0.119\pm 0.003\,\mbox{fm}^2$
\cite{Koes86}. Thus,  $\langle 0|({\bf r}_1-{\bf
R}_{c.m.})^2|0\rangle \,  > \, \langle 0|({\bf r}_3-{\bf
R}_{c.m.})^2|0 \rangle$.
 Due to the  symmetry of $H_{nr}$ with respect to the permutation
 ${\bf r}_1\leftrightarrow  {\bf r}_3$ for the case of equal quark masses,
 it is evident that   $(r^2_E)_n=0$
in the non-relativistic approximation. The relativistic correction
to
 $(r^2_E)_n$ comes from the correction to the center-of-mass vector and from
 the correction to the wave function:
 \begin{eqnarray}\label{delrn}
&& \delta (r^2_E)_n= \frac{2}{3}\langle 0| \left\{{\bf r}_1-{\bf
r}_3\, ,\,  \delta{\bf R}_0+ \delta{\bf R}_s\right\}
|0\rangle\, \nonumber\\
&& -\frac{2}{3}\langle 0|\left[({\bf r}_1-{\bf R})^2-
({\bf r}_3-{\bf R})^2\right]G_0H_B|0\rangle \nonumber \\
&&-\frac{2}{3}\langle 0|H_BG_0\left[({\bf r}_1-{\bf R})^2- ({\bf
r}_3-{\bf R})^2\right]|0\rangle\, ,
\end{eqnarray}
where $G_0$ is the non-relativistic reduced Green function
\begin{eqnarray}
G_0=[\varepsilon_0-H_{nr}+i0]^{-1}
(1-|0\rangle\langle 0|)\, ,
\end{eqnarray}
and $\varepsilon_0$ is the ground state binding energy in the
non-relativistic approximation.

Using the properties of $\delta{\bf R}_0$ and $\delta{\bf R}_s$
with respect to the transformations (\ref{subs}), it is easy to show
that the contribution to $\delta (r^2_E)_n$ (\ref{delrn}) from the
correction to the center-of-mass vector, as well as the
contribution of the spin-independent part of $H_B$, vanish. Due to
parity conservation the contribution of the part of $H_B$
being linear in spin is also zero. The spin-spin part $H^{(ss)}_B$ of the Breit
Hamiltonian  gives a  non-zero contribution to $\delta(r^2_E)_n$.
This operator exists only for the Lorentz-vector part $V_v(r)$ of
the potential:
\begin{eqnarray}
&&H^{(ss)}=\sum_{i\ne j} \frac{1}{3m_im_j}
 \triangle V_v({r}_{ij})\,({\bf s}_i\cdot{\bf s}_j) \nonumber \\
&&- \sum_{i\ne j} \frac{1}{6m_im_j}
[V_v^{\prime\prime}({r}_{ij})-V_v^{\prime}({r}_{ij})/{r}_{ij}]
[3({\bf s}_i\cdot{\bf \hat{r}}_{ij})({\bf s}_j\cdot{\bf
\hat{r}}_{ij})-{\bf s}_i\cdot{\bf s}_j]\,  ,
\end{eqnarray}
where ${\bf \hat{r}}_{ij}={\bf r}_{ij}/r_{ij}$. Substituting
$V_v(r)=-2\alpha_s/3r$ and averaging over the spin part of the
neutron wave function, we obtain
\begin{eqnarray}
&&H^{(ss)}=\frac{2\pi\alpha_s}{9\sqrt{2}m^2}\left[
\delta({\boldsymbol\rho})-2\delta({\boldsymbol\xi})-
2\delta({\boldsymbol\eta})\right] \, .
\end{eqnarray}
Using Jacobi variables, we can represent   $\delta (r^2_E)_n$ as
\begin{eqnarray}\label{delrn1}
&& \delta (r^2_E)_n= -\frac{1}{3}\langle
0|[\rho^2-\lambda^2]G_0H^{(ss)}|0\rangle -\frac{1}{3}\langle
0|H^{(ss)}G_0[\rho^2-\lambda^2] |0\rangle\, .
\end{eqnarray}
In order to calculate the matrix element (\ref{delrn1}) we follow
the prescription of \cite{Isgur00} and set $V(r)=Kr^2/2 $ in the
non-relativistic Hamiltonian $H_{nr}$. Then we obtain
\begin{eqnarray}\label{HnrJacobi1}
H_{nr}= \frac{ P^2}{2M}+\frac{ p^2_{\rho}}{2m}+ \frac{
p^2_{\lambda}}{2m}+\frac{3K(\rho^2+\lambda^2)}{2} \, .
\end{eqnarray}
Thus, we have two independent oscillators with equal frequencies
$\omega_\rho=\omega_\lambda=\sqrt{3K/m}$. Taking into account that
$(\rho^2-\lambda^2)|0\rangle$ is an  eigenfunction of $H_{nr}$
(\ref{HnrJacobi1}) with the excitation energy
$E-E_0=2\omega_\rho$, we have
\begin{eqnarray}\label{delrn2}
&& \delta (r^2_E)_n= \frac{1}{3\omega_\rho}\langle
0|[\rho^2-\lambda^2]H^{(ss)}|0\rangle=
 -\frac{\alpha_s}{3\sqrt{2\pi\omega_\rho m^3}}\, .
\end{eqnarray}
Thus, we obtain the negative value for $(r^2_E)_n$. The magnitude
of this quantity can also be made in agreement  with the experimental value by
taking the appropriate parameters.

\section{Electric polarizability}

Using the Jacobi  variables we obtain the operator of the internal
dipole moment ${\bf D}=e\sum e_i({\bf r}_i-{\bf R}_{c.m.})$ for
proton and neutron:
\begin{eqnarray}\label{dipole}
&& {\bf D}_p= e\left(
\sqrt{\frac23}{\boldsymbol\lambda}-\delta{\bf R}_{c.m.}\right)\,
,\quad {\bf D}_n= -e \sqrt{\frac23}{\boldsymbol\lambda}\, .
\end{eqnarray}
If we neglect $\delta{\bf R}_{c.m.}$ in ${\bf D}_p$, we
immediately obtain $(\alpha_0)_p=(\alpha_0)_n$ since $\alpha_0$ is
quadratic in ${\bf D}$ (see (\ref{alpha})). Therefore, the
difference between $(\alpha_0)_p$ and $(\alpha_0)_n$ arises only
due to the correction $\delta{\bf R}_{c.m.}$ in ${\bf D}_p$. The
contribution to this difference being  linear in $\delta{\bf
R}_{c.m.}$ reads:
\begin{eqnarray}\label{dalpha}
&&\delta\alpha= (\alpha_0)_p-(\alpha_0)_n=
e^2\left(\frac23\right)^{3/2} \left[\langle
0|{\boldsymbol\lambda}G\delta{\bf R}_{c.m.}|0\rangle + \langle
0|\delta{\bf R}_{c.m.}G{\boldsymbol\lambda}|0\rangle\right]\, ,
\end{eqnarray}
where $G$ is the reduced Green function accounting for the first
relativistic correction:
\begin{eqnarray}
G=[E_0-H_{nr}-H_B+i0]^{-1} (1-|0\rangle\langle 0|)\, .
\end{eqnarray}
If we replace $G$ by its non-relativistic limit $G_0$  and use the
symmetry with respect to the transformations (\ref{subs}), we
obtain zero as the result for $\delta\alpha$. Therefore, it is
necessary to take into account the corrections to the wave
function and to the Green function due to the spin-dependent part
of $H_B$. In addition, it is necessary to account for the term
quadratic in $\delta \textbf{R}_{c.m.}$ in (\ref{alpha}), and the
second-order relativistic correction to ${\bf R}_{c.m.}$. The
calculation of the latter is a very complicated problem. Since we
are  only going to demonstrate the possible cancellation between
$\delta\alpha$ and $\Delta\alpha$ (see (\ref{alpha})), we may
simplify our problem and assume, in a spirit of the diquark model
\cite{Isgur00}, that the relativistic effects reduce to the small
difference between the masses $m_1$ and $m_3$. As a result, we use
the non-relativistic Hamiltonian
\begin{eqnarray}\label{HnrJacobi2}
H_{nr}= \frac{ P^2}{2M}+\frac{ p^2_{\rho}}{2m_1}+ \frac{
p^2_{\lambda}}{2m_{\lambda}}+\frac{3K(\rho^2+\lambda^2)}{2} \,
\end{eqnarray}
and the expression (\ref{rcm0difm}) for $\delta{\bf R}_0$. Remind
that $M=2m_1+m_3$ and $m_{\lambda}=3m_1m_3/M$. Thus, the
corresponding frequencies are $\omega_\rho=\sqrt{3K/m_1}$ and
$\omega_\lambda=\sqrt{3K/m_\lambda}$. We can fix the parameters of
the model from the experimental values of $(r^2_E)_p$ and
$(r^2_E)_n$. Using (\ref{Jacobi}), (\ref{radii}) and
(\ref{HnrJacobi2}) it is easy to find that
 \begin{eqnarray}\label{radii1}
(r^2_E)_p&=&\frac{1}{m_1\omega_\rho}\left[1+\frac{(m_3^2-m_1^2)\omega_\rho}
{m_3M\omega_\lambda}\right]\approx \frac{1}{m_1\omega_\rho}\, , \nonumber \\
(r^2_E)_n&=&
-\frac{1}{2m_1\omega_\rho}\left[1-\frac{(2m_1-m_3)\omega_\rho}
{m_3\omega_\lambda}\right]\approx -\frac{5(1-x)}{6m_1\omega_\rho}
\, ,
\end{eqnarray}
where $x=m_1/m_3$. Substituting the experimental values of charge
radii into (\ref{radii1}), we obtain $x=0.8$ , and
$m_1\omega_\rho=6\cdot 10^4\,\mbox{MeV}^2$. Using the conventional
value $m_1=330\,$MeV, we have $\omega_\rho/m_1=0.6$. As a result
of simple calculations we have in our model:
\begin{eqnarray}
\Delta\alpha_p=\frac{e^2}{9m_1^2\omega_\rho}\quad ,\quad
\delta\alpha=-\frac{79e^2(1-x)}{108m_1^2\omega_\rho}+\frac{53e^2}{1296m_1^3}\,.
\end{eqnarray}
The second term in $\delta\alpha$ comes from the  correction being
quadratic in $\delta{\bf R}_{c.m.}$ and numerically is much
smaller than the first one. For the parameters of our model we
obtain
\begin{equation}
\frac{\delta\alpha+\Delta\alpha_p}{\Delta\alpha_p}=-0.1\, ,
\end{equation}
which leads to the approximate equality of the predicted proton
and neutron electric polarizabilities in agreement with the
experimental data.

\section*{Conclusion}

Thus we have demonstrated that the equality of the proton and
neutron electric polarizability can be explained in the frame of
constituent quark model. Therefore, it is not necessary to deal
with such \textit{ad hoc} contributions (for constituent quark
model) as mesonic currents. Though it has been shown in
\cite{lmsh01} that there are many sources of the relativistic
corrections of the same order as $\Delta\alpha$, due to the
isospin invariance it is crucial to take into account the
correction to the center-of-mass vector. If this correction is
neglected, the difference $(\alpha_\circ)_p-(\alpha_\circ)_n$
vanishes identically. The estimate for the relativistic
corrections made above with the use of a toy model serves only as
an illustration of the possibility to make the difference
$(\bar\alpha)_p-(\bar\alpha)_n$ consistent with the experimental
data at reasonable values of the parameters.

\section*{Acknowledgements}

One of us (A.I.M.) wishes to thank Zweites Physikalisches
Institut, University of G\"ottingen, for the very warm hospitality
during the stay when the part of this work has been done.
This work was supported by Deutsche Forschungsgemeinschaft through
contracts SCHU222 and 436 RUS 113/510.


\end{document}